\documentclass[runningheads]{meta}

\usepackage{epsfig}
\usepackage{algorithm}
\usepackage{algorithmic}
\usepackage{subfloat}
\usepackage{graphicx}
\usepackage{subfig} 
\usepackage{verbatim}

\usepackage{tikz}
\usepackage{pgfplots}

\begin{document}

\pagestyle{headings}

\mainmatter

\title{Obtaining the coefficients of a Vector Autoregression Model through minimization of parameter criteria}

\titlerunning{Vector Autoregression for Time Series}

\author{Alfonso L. Casta\~no\inst{1} \and Javier Cuenca\inst{2} \and Domingo Gim\'enez\inst{1} \and Jose J. L\'opez-Esp\'{\i}n\inst{3}\and Alberto P\'erez-Bernabeu\inst{4}}

\authorrunning{\ }

\institute{Department of Computing and Systems, University of Murcia, Spain.\\
\email{\{alfonsoluis.castanom,domingo\}@um.es}\\
\and
Department of Engineering and Technology of Computers, University of Murcia, Spain.\\
\email{jcuenca@um.es}\\
\and
Center of Operations Research, Miguel Hern\'andez University, Elche Campus,
 Spain.\\
\email{jlopez@umh.es}\\
\and
Department of Foundations of Economic Analysis, University of Alicante, Spain.\\
\email{apb74@alu.ua.es}
}

\maketitle

\section{Introduction}

VAR models \cite{Sims80} are a type of multi-equation model that linearly describe the simultaneous interactions and behaviour among a group of variables using only their own past. More specifically, a VAR is a model of simultaneous equations formed by a system of equations in which the contemporary values of model variables do not appear in any explanatory variable in the equations. The set of explanatory variables in each equation is a block consisting of lags of each of the model variables, and the block is the same for all the equations.

VAR models have been traditionally used in finance and econometrics
\cite{Coch05,Tsay05}.
With the arrival of Big Data, huge amounts of data are
being collected in numerous fields. 
Our group is studying the application of statistical models 
in health problems which, traditionally, have been applied in econometrics \cite{Lope16}.
To model time series we use Vector Autoregression Models (VAR). 
Tools exist to tackle this problem \cite{STATA}, but
the large amount of data, along with the availability of computational techniques and high performance systems,
advise an in-depth analysis of the computational aspects
of VAR, so large models can be solved efficiently with
today's computational systems.

To solve the model, Ordinary Least Squares (OLS) are used equation by equation. However, the practical challenge of its design lies in selecting the optimal length of the lag of the model. There are different strategies  to solve this problem \cite{luktkepohl2005,lutkepohl2009}: one of them is to examine some information criteria, for example Akaike (AIC) \cite{Akai73}, Schwarz (BIC) \cite{BIC} or Hannan-Quinn (HQC) \cite{Hannan1979} (these are the most well-known and used criteria but not the only ones).
This work aims to solve a VAR model by obtaining the coefficients through heuristic and metaheuristic algorithms, minimizing one parameter criterion, and also to compare with those coefficients obtained by OLS. 

\section{Computational aspects of VAR}

VAR are based on the idea that the value of a variable
at a time point depends linearly on the value of one or more variables at previous instants of
time. Since, most of the time, not all the variables will follow this pattern, there will be
dependent and independent variables. The latter will not be analyzed by the model, but will be used to
model dependent variables.

Let $y^{(j)}$ and $z^{(j)}$ denote the vector of dependent and independent variables at some time point
$j$. The vector $y^{(j)}$ depends on the value of $y$ and $z$ at $i$ and $k$ previous time points. The linear
dependence of $y^{(j)}$ with $y^{(j-t)}$ and $z^{(j-t)}$ is expressed by the matrices $A_t$ and $B_t$. Finally, $C$ 
denotes an independent vector to fit the model better. Thus, our VAR model has the form

$$
y^{(j)} \approx y^{(j-1)} A_1 + \ldots + y^{(j-i)} A_i + z^{(j-1)} B_1 + \ldots + z^{(j-k)} B_k + C
$$

\noindent Since matrices $A$, $B$ and $C$ are the unknown terms that we want to ascertain, we can build the system

\begin{equation}
\left(
\begin{array}{c}
y^{(i)} \\
\vdots \\
y^{(t)}
\end{array}
\right)
\approx
\left(
\begin{array}{ccccccc}
y^{(i-1)} & \ldots &y^{(0)} & z^{(i-1)} &\ldots &z^{(i-k)} &1\\ 
& \vdots & & &\vdots & &\vdots \\
y^{(t-1)} & \ldots &y^{(t-i)} & z^{(t-1)} &\ldots &z^{(t-k)} &1
\end{array}
\right)
\left( 
\begin{array}{c}
A_1 \\
\vdots \\ 
A_i \\
B_1 \\
\vdots \\ 
B_k \\
C
\end{array}
\right)
\label{ecu:VAR}
\end{equation}

\noindent In this way, the solution of the system is traditionally approached by Ordinary Least Squares (OLS), maximum likelihood, etc. Here, an approach using heuristic and metaheuristic algorithms (with special emphasis on Genetic Algorithms \cite{Holland1973}, Scatter Search \cite{Glover2003}, GRASP \cite{Resende2003} and Tabu Search \cite{Glover1997} and combinations of them) for minimizing some parameter criteria is studied and compared with traditional methods.

We plan to study some computational problems arising from this approach:

\begin{itemize}

\item The matrix formulation allows us to apply matrix computations \cite{GoVa13}. QR or LQ decompositions can be applied to simplify the system, so reducing the
time required to solve the system, but the Toeplitz-type structure in equation \ref{ecu:VAR} advises the
adaptation of algorithms for structured matrices. 

\item A new problem occurs when we do not know how many previous time points the variables depend
on or which variables are dependent and independent. Finding the best model configuration is
a combinatorial optimization problem with a huge computational cost. Therefore, an exhaustive
search is not suitable and other techniques should be applied.

\item Since the outlined approach has a big computational cost, we propose two levels of parallelism to
enhance performance.
On the one hand, high performance linear algebra
subroutines based on BLAS and LAPACK can be used, so the
parallelism is intrinsically exploited. 
On the other hand, we consider the use of explicit parallelism to simultaneously compute independent
model configurations.

\end{itemize}

\noindent {\bf Acknowledgements}.
This work was supported by the Spanish MINECO, as well as European Commission FEDER funds, under grants TIN2016-80565 and TIN2015-66972-C5-3-R.

\bibliographystyle{abbrv}
\bibliography{bibliografia.bib}  

\begin{thebibliography}{10}

\bibitem{Akai73}
H.~Akaike.
\newblock Information theory and an extension of the maximum likelihood
  principle.
\newblock In {\em Proc. 2nd Int. Symp. on Information Theory}, pages 267--281,
  1973.

\bibitem{Coch05}
J.~H. Cochrane.
\newblock Time {S}eries for {M}acroeconomics and {F}inance.
\newblock Graduate School of Business, University of Chicago, 2005.

\bibitem{Glover2003}
F.~Glover and G.~A. Kochenberger.
\newblock Handbook of metaheuristics.
\newblock {\em Kluwer Academic}, 2003.

\bibitem{Glover1997}
F.~Glover and M.~Laguna.
\newblock Tabu {S}earch.
\newblock {\em Kluwer Academic}, 1997.

\bibitem{GoVa13}
G.~Golub and C.~F.~V. Loan.
\newblock {\em Matrix Computations}.
\newblock The John Hopkins University Press, fourth edition, 2013.

\bibitem{Hannan1979}
E.~J. Hannan and B.~G. Quinn.
\newblock The determination of the order of an autoregression.
\newblock {\em Journal of the Royal Statistical Society: Series B
  (Methodological)}, 41:190--195, 1979.

\bibitem{Holland1973}
J.~H. Holland.
\newblock {GA} and the optimal allocation of trials.
\newblock {\em SIAM Journal on Computing}, 2(88–105), 1973.

\bibitem{Lope16}
J.~J. L\'opez-Esp\'{\i}n.
\newblock Development and analysis of algorithm for the best econometric model
  in health problems (in {S}panish).
\newblock University Miguel Hern\'andez of Elche, 2016.

\bibitem{luktkepohl2005}
H.~L{\"u}tkepohl.
\newblock {\em New introduction to multiple time series analysis}.
\newblock Springer Science \& Business Media, 2005.

\bibitem{lutkepohl2009}
H.~L{\"u}tkepohl.
\newblock Econometric analysis with vector autoregressive models.
\newblock {\em Handbook of Computational Econometrics}, pages 281--319, 2009.

\bibitem{Resende2003}
M.~G.~C. Resende and C.~C. Ribeiro.
\newblock Greedy randomized adpative search procedures.
\newblock {\em Kluwer Academic}, 2003.

\bibitem{BIC}
G.~Schwarz.
\newblock Estimating the dimension of a model.
\newblock {\em Ann. Statist}, 6:461--464, 1978.

\bibitem{Sims80}
C.~A. Sims.
\newblock Macroeconomics and reality.
\newblock {\em Econometrica}, 48(1):1--48, 1980.

\bibitem{STATA}
{Stata: Data Analysis and Statistical Software}.
\newblock {\tt https://www.stata.com/}.

\bibitem{Tsay05}
R.~S. Tsay.
\newblock {\em Analysys of {F}inancial {T}ime {S}eries}.
\newblock John Wiley \& Sons, second edition, 2005.

\end{thebibliography}

\end{document}